\documentclass[manuscript]{aastex}
\shorttitle{Detection of Substructure in MG0414+0534}
\begin{document}
\title{Detection of Substructure in the Gravitationally Lensed Quasar MG0414+0534 using Mid-Infrared and Radio VLBI Observations}
\author{Chelsea L. MacLeod\altaffilmark{1}, Ramsey Jones\altaffilmark{2}, 
  Eric Agol\altaffilmark{2}, and Christopher S. Kochanek\altaffilmark{3}}
\altaffiltext{1}{Physics Department, United States Naval Academy, Annapolis, MD
  21403; macleod@usna.edu}
\altaffiltext{2}{Astronomy Department, University of Washington, Seattle, WA
  98195}
\altaffiltext{3}{Department of Astronomy and the Center for Cosmology and
  Astroparticle Physics, The Ohio State University, Columbus, OH
  43210}

\begin{abstract}
We present 11.2 $\mu$m observations of the gravitationally lensed, 
radio-loud $z_s=2.64$ quasar MG0414+0534, obtained using the Michelle camera on Gemini North. 
We find a flux ratio anomaly of $\rm{A2/A1}= 0.93 \pm 0.02$ for the quasar images A1 and A2. 
When combined with the 11.7 $\mu$m measurements from Minezaki et al.\ (2009), the $\rm{A2/A1}$ 
flux ratio is nearly $5 \sigma$ from the expected ratio for a model based on the two visible lens galaxies. 
The mid-IR flux ratio anomaly can be explained by a satellite (substructure), 0\farcs3 
Northeast of image A2, as can the detailed VLBI structures of the jet produced by the quasar. 
When we combine the mid-IR flux ratios with high-resolution VLBI measurements, we find a 
best-fit mass between $10^{6.2}$ and $10^{7.5}\;M_\odot$ inside the Einstein radius
for a satellite substructure modeled as a singular isothermal sphere at the 
redshift of the main lens ($z_l=0.96$).  We are unable to set an interesting limit 
on the mass to light ratio due to its proximity to the quasar image A2.
While the observations used here were technically difficult, surveys of flux anomalies 
in gravitational lenses with the $James~Webb~Space~Telescope$ will be simple, 
fast, and should well constrain the abundance of substructure in dark matter haloes. 
\end{abstract}
\maketitle

\section{Introduction}

The missing satellite problem, where cosmological simulations of cold dark matter (CDM) predict a significantly larger fraction of lower mass satellites around galaxies than is detected, represents a major puzzle in the study of structure formation \citep{kly99,moo99}. In the case of the Milky Way, the Sloan Digital Sky Survey has steadily found additional, faint satellites, but the total numbers are still far lower than the expected abundances of subhaloes \citep[e.g.,][]{wil05,bel07}.  The simplest solution in the context of CDM models is to suppress star formation in low-mass satellites, probably through heating and baryonic mass loss as the universe re-ionizes to leave a population of dark satellites \citep[e.g.,][]{kly99,bul00} with assistance from the relative streaming of baryons and dark matter after recombination \citep{hirata}. Locally there is some hope of finding these dark satellites through $\gamma$-rays emitted by dark matter annihilation, but the likelihood of detection depends heavily on the properties of dark matter \citep[e.g.,][]{str08}. 

Gravitational lensing offers an alternate method to detect subhaloes.
In some strong lenses (e.g.\ MG0414+0534, \citealt{ros00}; MG2016+112,
\citealt{mey06}; HE0435--1223, \citealt{koc06}, \citealt{fad12};
B1938+666, \citealt{veg12}), satellite galaxies of the primary lens
can be detected astrometrically through their effects on the image positions.
Potentially, the most massive satellites can also be detected through
their effects on time delay ratios \citep{kee09}. But as emphasized by
\citet{mao98} and \citet{met02}, image fluxes are sensitive to
perturbations even from very low-mass satellites ($\lesssim 10^6 M_{\odot}$).
Moreover, we observe many lenses with ``flux ratio anomalies'' where the
relative image brightnesses cannot be explained by simple, central
lens galaxies \citep[see][]{ eva03,koc04,con05,yoo06a,yoo06,kra11}.
While current results are consistent with the expectations of CDM 
\citep{dal02,veg12}, there are still too few lenses to precisely test 
the CDM model using flux ratio anomalies.  For this technique, the challenge 
is that image fluxes can also be affected by extinction and 
microlensing in the primary lens at the most easily observed wavelengths, 
although the wavelength dependencies of these three
effects can help to disentangle them from each other
\citep[e.g.,][]{ago09,mun11}.  The millilensing signal by 
substructures is best isolated at radio or rest-frame mid-infrared (mid-IR, 
$\sim$4--100~$\mu$m) wavelengths, where both extinction and microlensing 
effects from the lens galaxy are negligible (stars can only magnify sources 
smaller than $\sim$0.1~pc). 
Ideally, our observations (3.1~$\mu$m in the rest frame) would be at still longer wavelengths, but simulations indicate that microlensing effects should be $\lesssim 0.1$~mag for our observations \citep{mar12,slu13}.
The mid-IR images are also unaffected by interstellar scattering, 
or scintillation, which can further perturb the radio images, although few lenses show the strong radio wavelength-dependent signatures expected from these effects \citep{koc04}.  The biggest problem for using radio wavelengths is simply that most quasars are radio-quiet, leaving only a handful of objects to be studied.  While mid-IR observations are challenging, they are ideal for searching for substructures in lens galaxies.  

Here we present mid-IR observations of the quadruply lensed, radio-loud $z_s=2.64$ quasar MG0414+0534, obtained using the Michelle camera on Gemini North.  This system is known to be affected by extinction and microlensing in the optical and near-IR \citep{hew92,law95,fal97,sch02,bat08,bat11,poo12}, but the flux ratio anomaly persists into the mid-IR and radio. Our new measurements have slightly higher precision than those by \cite{min09} and we then consider more detailed models.  This work is an extension of the study in \cite{mac09}, where a companion lens galaxy was detected through the mid-IR flux ratios of the four-image lens H1413+117.  Here, we show that the flux ratio anomaly in MG0414+0534 indicates the presence of low-mass substructure in the lens based on models of the individual or combined radio and mid-IR observations of this lens.   Our data are described in Section~\ref{sec:obs}, our lens models are presented in Section~\ref{sec:mod}, and our conclusions are in Section~\ref{sec:disc}.

\section{Flux measurements and errors}
\label{sec:obs}
\begin{figure}[h!]
\centerline{
\includegraphics[width=6in,viewport=0 0 510 166,clip]{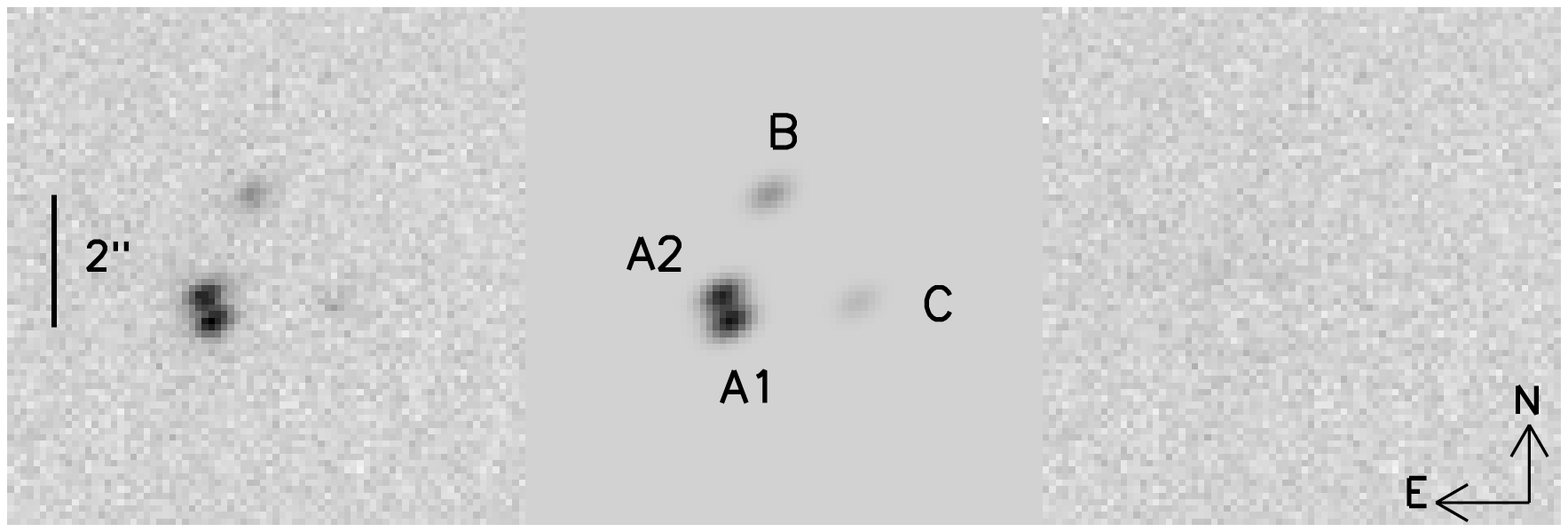}}
\centerline{\includegraphics[width=6in,viewport=0 0 510 166,clip]{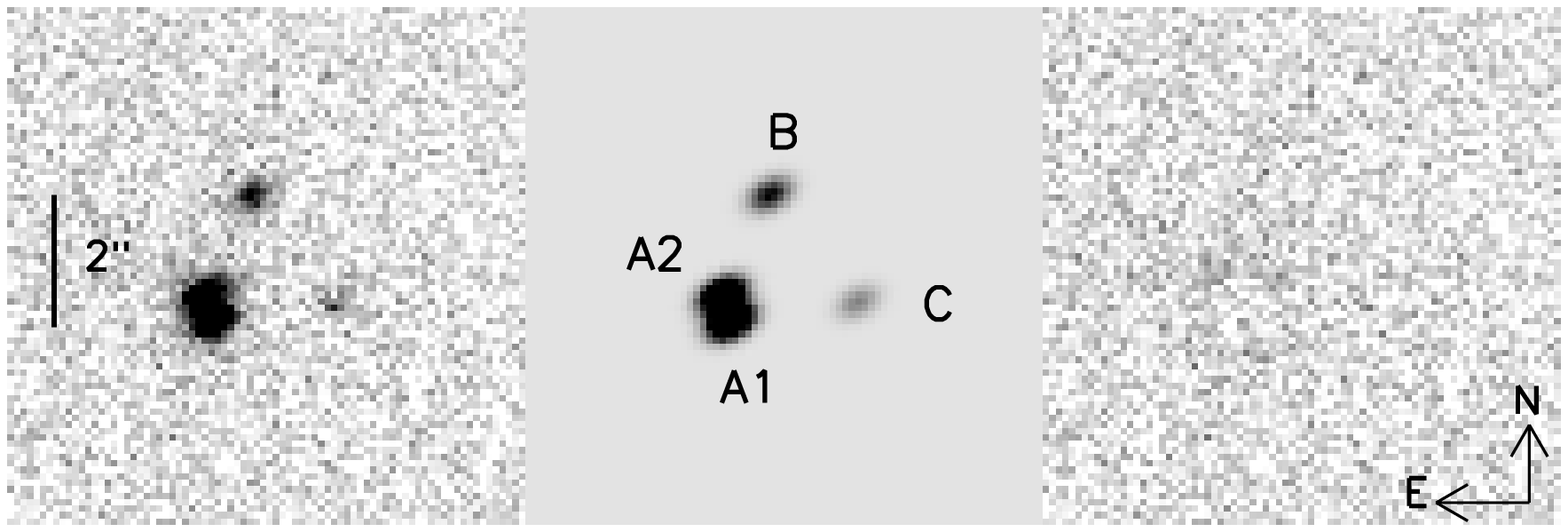}}
\caption{\footnotesize The left, middle, and right panels show the 11.2~$\mu$m
  data, the best-fit model based on 2-D
  Gaussians, and the model residuals, respectively, for MG0414+0534. 
  In the bottom panels, the images have been scaled to increase the visibility of B and C. }
\label{fig:images}
\end{figure}

We observed MG0414+0534 with the Michelle camera \citep{roc04} on Gemini North at
11.2~$\mu$m (F112B21 filter) on 2 Jan 2006. The observing time was 1 hour, of
which 940.8 seconds were spent on source. The data were processed with the standard Gemini
pipeline {\it mireduce}. Our analysis starts with the
coadded chop and nod subtracted image. These initial images have 16
vertical stripes due to the 16 readout channels.  We
corrected for the stripes by subtracting the median of each stripe after masking
the region containing the quasar images. This procedure also removes
any residual sky flux.  Figure~\ref{fig:images} shows the resulting image. 

We fit the four lensed images using a 2-D Gaussian to model the PSF,
fixing the relative image positions to those measured from the $Hubble~Space~Telescope$ 
($HST$) images\footnote{http://www.cfa.harvard.edu/castles/} \citep{fal97}. 
Since the sky is so much brighter than 
the quasar images, we use the standard deviation of the background
pixels as an estimate of the errors in the flux of each pixel.  The model has nine free
parameters:  the fluxes of the four images (4), the position of image B (2), the
standard deviation of the Gaussian PSF along the two principal directions (2),
and the angle of the major axis.  We found the best-fit model using
non-linear Levenberg-Marquardt optimization.  We used a Markov chain
Monte Carlo (MCMC) with 10$^4$ steps to compute the parameter
uncertainties, and ran multiple
chains to check the convergence of the best-fit parameters and their
uncertainties.  We also estimated the errors by simply computing the
goodness of fit as a function of the flux of each image, marginalizing over
all other variables.  

\begin{figure}[h!]
    \epsscale{0.7}
    \plotone{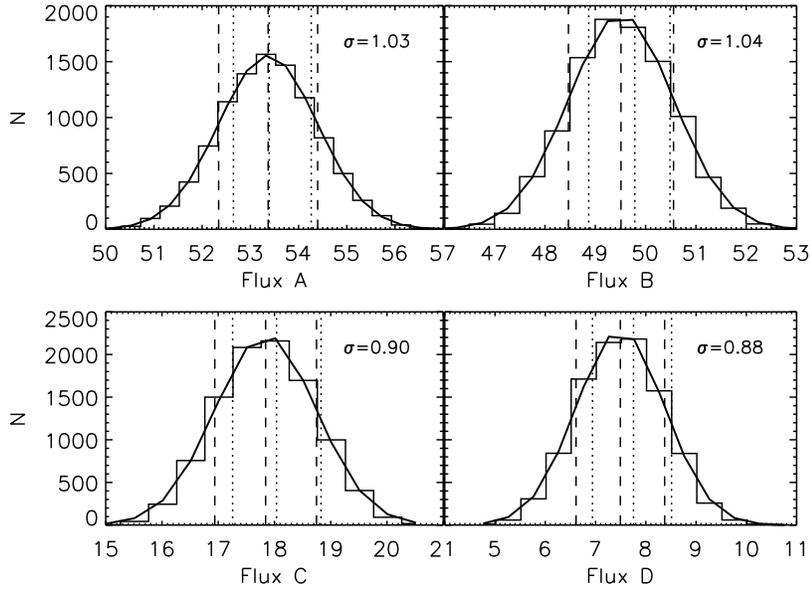}
\caption{\footnotesize MCMC probability distributions for the 
  MG0414+0534 image fluxes  (in arbitrary units).
 The vertical dashed lines show the mean and $1\sigma$ error bars determined from the
  best-fit Gaussian curve.
  These are compared to the best-fit flux value and 68\% confidence limits (or parameter values at which
  $\Delta\chi^2 = 1$ from the minimum $\chi^2$) that result from the $\chi^2$ minimization
  technique (see text), which are shown by vertical dotted lines.
  The Gaussian widths ($\sigma$) are listed in the top right corners. 
  For the uncertainties reported in Table~\ref{tab:errors}, we instead use the dispersion of flux $ratios$ from the MCMC. }
\label{fig:MG0414errors}
\end{figure}

Figure~\ref{fig:images} shows the best-fit model and its residuals,
and Figure~\ref{fig:MG0414errors} shows the MCMC and $\Delta\chi^2$
estimate for the fluxes (in arbitrary units) and their uncertainties. 
In order to include any covariances between the image fluxes, we use the actual distribution of flux ratios 
from the MCMC approach to estimate the flux ratio uncertainties. 
We calibrated the flux measurements using observations of the standard star 
Rho Orionis (HR 1698) and the \citet{coh99} flux calibration 
at 11.2~$\mu$m (the interpolated flux is $\simeq 6.43$~Jy). 
We obtained a total flux of $33.4 \pm 1.3$(statistical)~$\pm 3.2$(systematic)~mJy 
at 11.2~$\mu$m from aperture photometry, which is lower than the value of 
$39.2 \pm 1.4$~mJy in \citet{min09}. This discrepancy may be due to the fact that 
our filter has a shorter central wavelength and that the SED is rising towards longer
wavelengths. The resulting mid-IR flux ratios are
listed in Table~\ref{tab:errors} along with the mid-IR results from \citet{min09}, 
and the combined results. 
The results agree given their mutual uncertainties and show that the anomaly, 
while not as strong as in the near-IR \citep{bat08}, persists into the mid-IR.   
This wavelength behavior is a first indication that the anomalous flux ratio 
in the near-IR is not produced solely by microlensing, as assumed in \citet{bat08}, 
but also by substructure in the lens. 

\begin{deluxetable}{c c c c}
\tablecolumns{4}
\tablewidth{0pt}
\tablecaption{Relative Mid-IR Fluxes for the Images of MG0414+0534 \label{tab:errors}}
\tablehead{
  \colhead{Images} \vspace{-0.2cm} & \multicolumn{3}{c}{Flux Ratio} \\
   \vspace{-0.2cm}  & \colhead{This work} & \colhead{Minezaki et al.\ (2009)} & \colhead{Combined}}
\startdata
 \vspace{-0.2cm} A2/A1 & 0.926 $\pm$ 0.025 & 0.90 $\pm$ 0.04 & 0.919 $\pm$ 0.021\\ 
  \vspace{-0.2cm} B/A1 & 0.338 $\pm$ 0.017 & 0.36 $\pm$ 0.02 & 0.347 $\pm$ 0.013\\
                  C/A1 & 0.145 $\pm$ 0.016 & 0.12 $\pm$ 0.03 & 0.139 $\pm$ 0.014\\
\tableline
\enddata
\end{deluxetable}


\clearpage

\section{Models}
\label{sec:mod}
In a multiple-image strong gravitational lens, a close pair of bright images such as A1 and A2 results if the quasar is near a fold caustic created by the main lens galaxy.  Their flux ratio is expected to be unity for a smooth lens model \citep[e.g.,][]{kee05}, whereas we measure a flux ratio of A2/A1~$=0.93\pm 0.02$ for MG0414+0534.  Since we are insensitive to extinction or microlensing effects created by the lens galaxy in the mid-IR, the anomalous ratio must be attributed to millilensing by substructure in the lens, as was also argued by \cite{min09}.  

We model the system with 
LENSMODEL\footnote{http://redfive.rutgers.edu/$\sim$keeton/gravlens/}
\citep{kee01}, using the combined mid-IR flux ratios listed in Table~\ref{tab:errors} (last column) as 
constraints on our models.  We use the measured $HST$ WFPC/WFPC2
positions of the main, central lens galaxy, the secondary lens 
``X'' \citep{sch93,ros00}, and the image positions from the CASTLES database \citep{fal97}.  
The component positions are listed in Table~\ref{tab:pos}.  The main lens galaxy (G1) is 
modeled as a singular isothermal ellipsoid (SIE) combined with an external
shear, as this is the simplest plausible model for a lens.  There is
considerable evidence in favor of the isothermal profile, and the 
properties of 4-image systems other than time delays are very
insensitive to the radial mass distribution \citep[see][]{mey06,koo09}.  
We include weak priors on the ellipticity of the main lens 
($1-b/a=0\pm 1$), and the external shear ($\gamma=0.05\pm 0.05$).
The expectation value of 0 for the ellipticity is based on the relatively 
round isophotes derived from the $HST$ photometry of the lens \citep{fal97,kee97}. 
\citet{ros00} found that the presence of an observed nearby galaxy, object X \citep{sch93}, 
hereafter ``G2'', in their lens models was essential for reproducing the image astrometry.
For G2, we tried two different models: a singular isothermal sphere (SIS) or a 
pseudo-Jaffe (pJaffe) profile with a scale length $a'$ that is allowed to vary.

\begin{deluxetable}{c c c c}
\tablecolumns{4}
\tablewidth{0pt}
\tablecaption{MG0414+0534 Image and Lens Positions \label{tab:pos}}
\tablehead{{\bf MG0414} component & $HST$ position ('') & Rotated $0.1^{\circ}$ ('') & Error ('')}
\startdata
A1    & $(-0.600,-1.942)$ & $(-0.596,-1.943)$& 0.003 \\  
A2    & $(-0.732,-1.549)$ & $(-0.729,-1.550)$& 0.003 \\
 B    & $(0.000,0.000)  $ & $(0.000,0.000)$  & 0.003 \\
 C    & $(1.342,-1.650) $ & $(1.345,-1.648)$ & 0.003 \\
G1    & $(0.472,-1.277) $ & $(0.474,-1.276)$ & 0.003 \\
G2 (=X)& $(0.857,0.180) $ & $(0.857,0.181)$  & 0.011 \\
\tableline
\enddata
\tablecomments{ The positions are taken from the CASTLES database \citep{fal97} and are relative to image B, 
  where negative R.A.\ values are eastward of image B. 
  We rotate the positions counter-clockwise by $0.1^{\circ}$ 
  when adding the radio constraints to our lens models. }
\end{deluxetable}

Using the mid-IR flux ratios as constraints, we obtain the best-fit two--galaxy 
models described in Table~\ref{tab:modelsMG0414}. The models have 17
total constraints (the positions of 4 lensed images and both lens galaxies,
the 3 flux ratios, plus the priors on the ellipticity
$e$ and shear $\gamma$) and 12 free parameters (the mass scale $b_1$,
position, ellipticity $e$ and position angle $\theta_e$ for G1, the
amplitude $\gamma$ and position angle $\theta_{\gamma}$ of the external shear, the mass
scale $b_2$ and position for G2, and the source position), plus one more free parameter for the scale 
length $a'_2$ when using a pseudo-Jaffe profile.  In the Table, we describe our 
best two--galaxy models when using SIE+SIS or SIE+pJaffe profiles.   
We find a slightly better fit when using an SIS profile for G2, with  
$\chi^2=31.1$ for dof=5 degrees of freedom and a best-fit flux
ratio $A2/A1=1.01$.    The
discrepancy between the observed and predicted mid-IR flux 
ratios in our two--galaxy models suggests the presence of substructure
along the line of sight or associated with the main lens galaxy at nearly 
a $5\sigma$ level.

\begin{deluxetable}{c | c c c | c c}
\tablecolumns{6}
\tablewidth{0pt}
\tabletypesize{\scriptsize}
\tablecaption{MG0414+0534 Modeling Results \label{tab:modelsMG0414}}
\tablehead{	               & G1+G2(pJaffe)                    & G1+G2                          & G1+G2+G3              & G1+G2                 & G1+G2+G3 \\
\tableline
       & \multicolumn{3}{|c|}{{\bf (Mid-IR only)}}  & \multicolumn{2}{|c}{\bf (Mid-IR + radio)} }
\startdata
$\chi^2/$dof   	               & 28.8/4 	                  & 31.1/5                         & 6.11/2                & 103/41                & 59.5/38                    \\ 
$\chi^2_{pos}$                  &  4.83$\phantom{00}$              & $\phantom{0}$5.46$\phantom{0}$ & 0.142                 & 8.45$\phantom{0}$    & 5.51                       \\
$\chi^2_{flux}$                 &  22.7$\phantom{00}$              & 23.9$\phantom{00}$             & 4.71$\phantom{0}$     & 90.5$\phantom{00}$    & 52.1$\phantom{00}$         \\
$\chi^2_{gal}$                  &  0.746                           & 1.18$\phantom{0}$              & 0.539                 & 3.17$\phantom{00}$    & 1.06$\phantom{00}$         \\
$\chi^2_{prior}$                &  0.613      	                  & 0.634                          & 0.720                 & 0.928                 & 0.849                      \\
$b_1$                          &  1\farcs104                      & 1\farcs068                     & 1\farcs091            & 1\farcs083            & 1\farcs084                 \\
$e$                            &  0.298                           & 0.302                          & 0.224                 & 0.242                 & 0.238                      \\
$\theta_e~(^{\circ}\rm{E~of~N})$&  $-88.5\phantom{00}$             & $-88.0\phantom{00}$            & $-85.4\phantom{00}$   & $-84.5\phantom{00}$   & $-83.9\phantom{00}$        \\
$\gamma$                       & 0.086                            & 0.087                          & 0.091                 & 0.097                 & 0.094                      \\
$\theta_{\gamma}~(^{\circ}\rm{E~of~N})$& 47.5$\phantom{00}$         & 47.2                           & 54.7                  & 53.6$\phantom{00}$    & 53.1$\phantom{00}$         \\
$b_2$                          & 0\farcs199	                  & 0\farcs193                     & 0\farcs170            & 0\farcs185            & 0\farcs176                 \\
$b_3$		               & ...                              & ...                            & 0\farcs0028           & ...                   &0\farcs007 \\
($x_3$,$y_3$)		       & ...                              & ...                            & ($-0\farcs91$, $-1\farcs56$) & ...             & ($-0\farcs97$, $-1\farcs39$) \\
Mid-IR Flux Ratios\\                                                                                                                                                           
\hline                                                                                                                                                                         
A2/A1		               & 1.0018                           & 1.0058                         & 0.9218                & 1.0320                & 0.9227                      \\
B/A1		               & 0.3318                           & 0.3415                         & 0.3438                & 0.3308                & 0.3354                      \\
C/A1		               & 0.1690                           & 0.1712                         & 0.1691                & 0.1614                & 0.1680                      \\
{\bf Source Positions}\\                                                                                                                                                           
\hline                                                                                                                                                                         
Mid-IR		               &  (0\farcs388, -1\farcs068) & (0\farcs404, -1\farcs026) & (0\farcs394, -1\farcs053) & (0\farcs404, -1\farcs039)         & (0\farcs397, -1\farcs045)       \\
Radio a		               & ...                           & ...                         & ...                & (0\farcs402, -1\farcs040)         & (0\farcs395, -1\farcs046)       \\
Radio b		               & ...                           & ...                         & ...                & (0\farcs403, -1\farcs040)         & (0\farcs396, -1\farcs045)       \\
Radio c		               & ...                           & ...                         & ...                & (0\farcs415, -1\farcs033)         & (0\farcs408, -1\farcs039)       \\
Radio d		               & ...                           & ...                         & ...                & (0\farcs392, -1\farcs044)         & (0\farcs385, -1\farcs050)       \\
\enddata
\tablecomments{\scriptsize The best-fit parameters for our two-- and three--galaxy models based on the mid-IR data 
  alone (left three columns) and after adding the VLBI constraints (right two columns).
  The subscripts differentiate between the main lens (1), secondary lens galaxy 
  G2/object ``X'' (2), and a third lens mass (3). 
  In the first column, G2 is modeled as a pseudo-Jaffe potential with 
  scale length $a'_2=2\farcs53$. In the remaining columns, G2 is modeled as an SIS.  
  The $\chi^2$ has
  contributions from the image positions ($\chi^2_{pos}$), flux ratios ($\chi^2_{flux}$),
  lens position ($\chi^2_{gal}$), and the priors on $e$ and $\gamma$ ($\chi^2_{prior}$).}
\end{deluxetable}

\begin{figure}[h!]
  \centering
  \epsscale{.4}
  \plotone{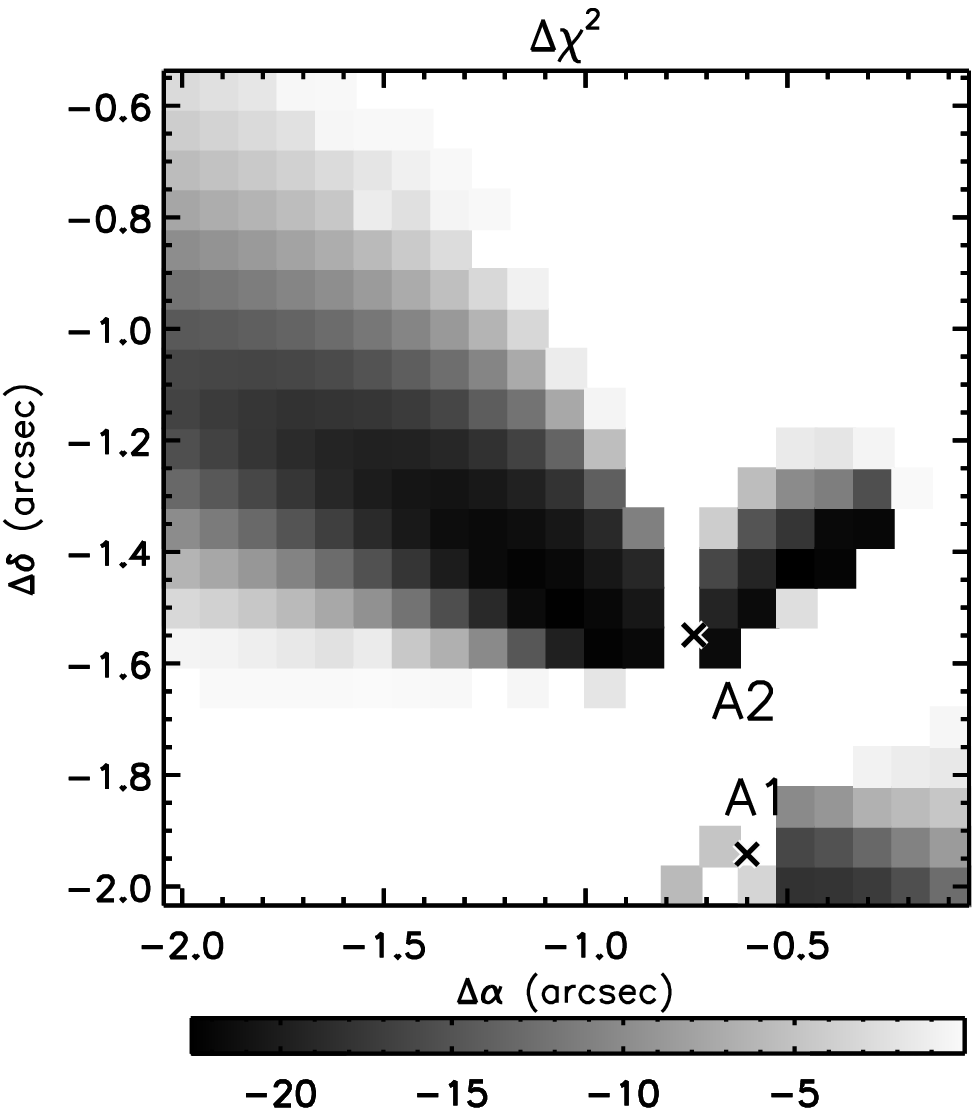}
  \plotone{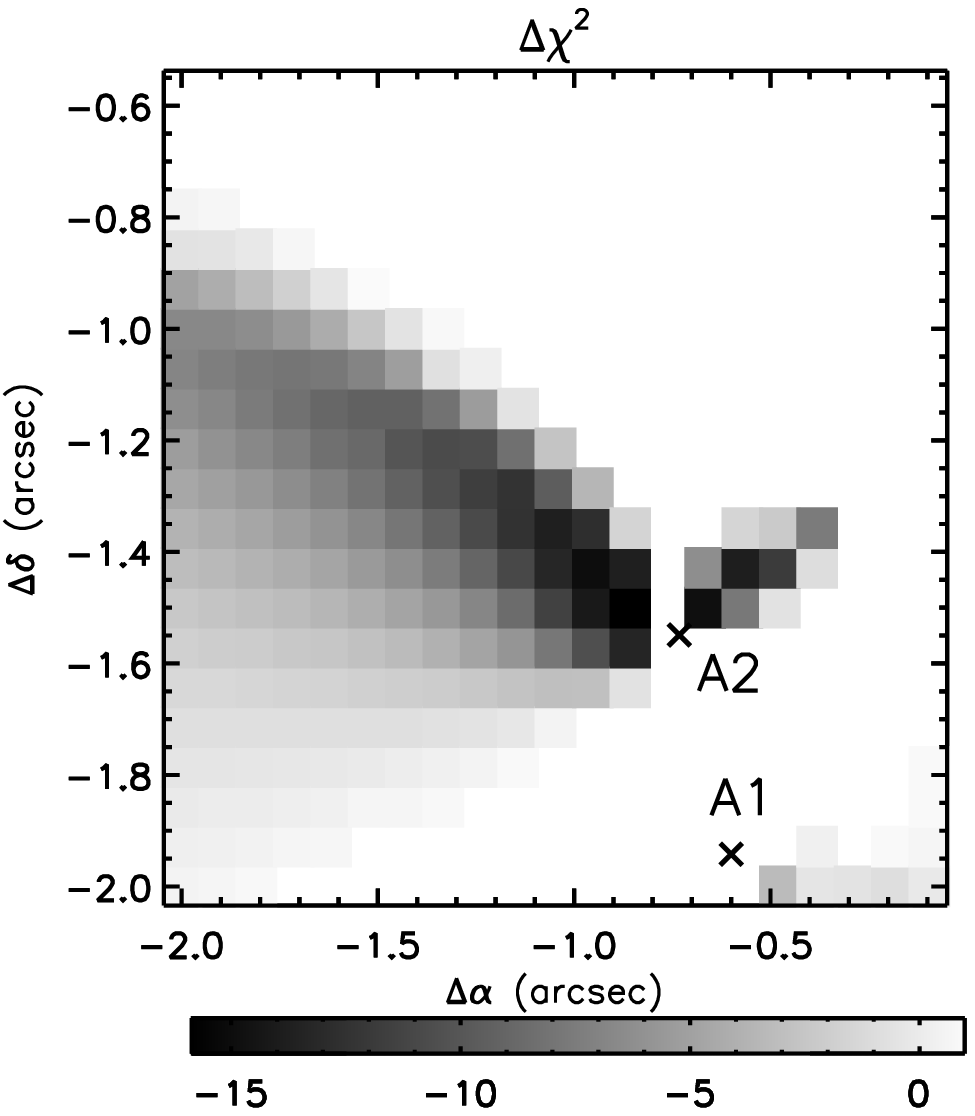}
  \caption{\footnotesize $\Delta\chi^2$ improvement resulting from adding
    a third lens galaxy to MG0414+0534 as a function of its 
    position. In the left panel, only the mid-IR constraints are used in the models. 
    In the right panel,  only the radio constraints are used.   
      The white regions are where the
    $\chi^2$ either matched or exceeded the best value from the
    2-galaxy models.  North is up, East is left.}
  \label{fig:3gal}
\end{figure}

We modeled the substructure by adding a third lens galaxy (G3) with an
SIS profile at various positions surrounding the lens and 
searched for a better fit to the observed flux ratios. Here, we adopt 
an SIS rather than a pseudo-Jaffe profile for G2 since the former 
yielded a better fit among the two--galaxy models.    At 
each position in the grid, the mass scales $b_1$, $b_2$, and $b_3$
for galaxies G1, G2, and G3 were allowed to vary, as well as the
ellipticity of G1 and the external shear,
until a best-fit three--galaxy model was obtained. Here, dof=6, since
we are holding the positions for G1 and G2 fixed for each model. 
Figure~\ref{fig:3gal} (left panel) shows the resulting
$\Delta\chi^2$ (relative to the two--galaxy model). It can be seen that the addition of substructure
improves the two--galaxy model when placed Northeast or Northwest of image A2, 
although the range of positions that lead to a similar improvement is more 
limited Northwest of A2. 
We consider the region Northeast of A2 as an approximate location for G3 in 
our initial three--galaxy model since it ultimately yielded lower $\chi^2$ 
values than the Northwest region.  We obtain the best-fit parameters in 
Table~\ref{tab:modelsMG0414}, where the coordinates of G3 are treated as free parameters. 
We find that an SIS with $b_3\approx 0\farcs0028$ located at $(x_3,y_3)=(-0\farcs91,-1\farcs56)$
leads to the best improvement, with $\Delta\chi^2= 25$ ($\chi^2/$dof~$=3$), although as 
described in Section~\ref{sec:vlbi}, a similar fit can be achieved if $b_3$ is larger and
G3 is slightly farther away from image A2.  
Assuming the measured redshift of $z_l=0.96$ for G1 \citep{ton99}, 
we estimate a best-fit mass of $\sim 3\times 10^6\;M_\odot$ and $\sim 10^{10}\;M_\odot$ 
enclosed within the Einstein radii of G3 and G2, respectively, 
based on the mid-IR constraints alone, as compared to $\sim 5\times 10^{11}\;M_\odot$ for G1.  
 The statistical significance of finding $\Delta\chi^2=25$ for 3 
additional parameters is 0.28 (a 72\% confidence level) based on the F-test.  
When G2 is modeled as a pseudo-Jaffe potential, we find similar results for a 
three--galaxy model but with $\chi^2/$dof~$=5.8$ (and an F-test confidence level of 45\%).  
 We conclude that G2 is better modeled as an SIS than a pseudo-Jaffe 
potential and therefore do not list the model parameters for a 
three--galaxy (SIE+pJaffe+SIS) model here.  
 
\subsection{Radio Constraints: 8.4 GHz VLBI Structures}
\label{sec:vlbi}

\begin{figure}[h]
  \epsscale{.9}
  \plotone{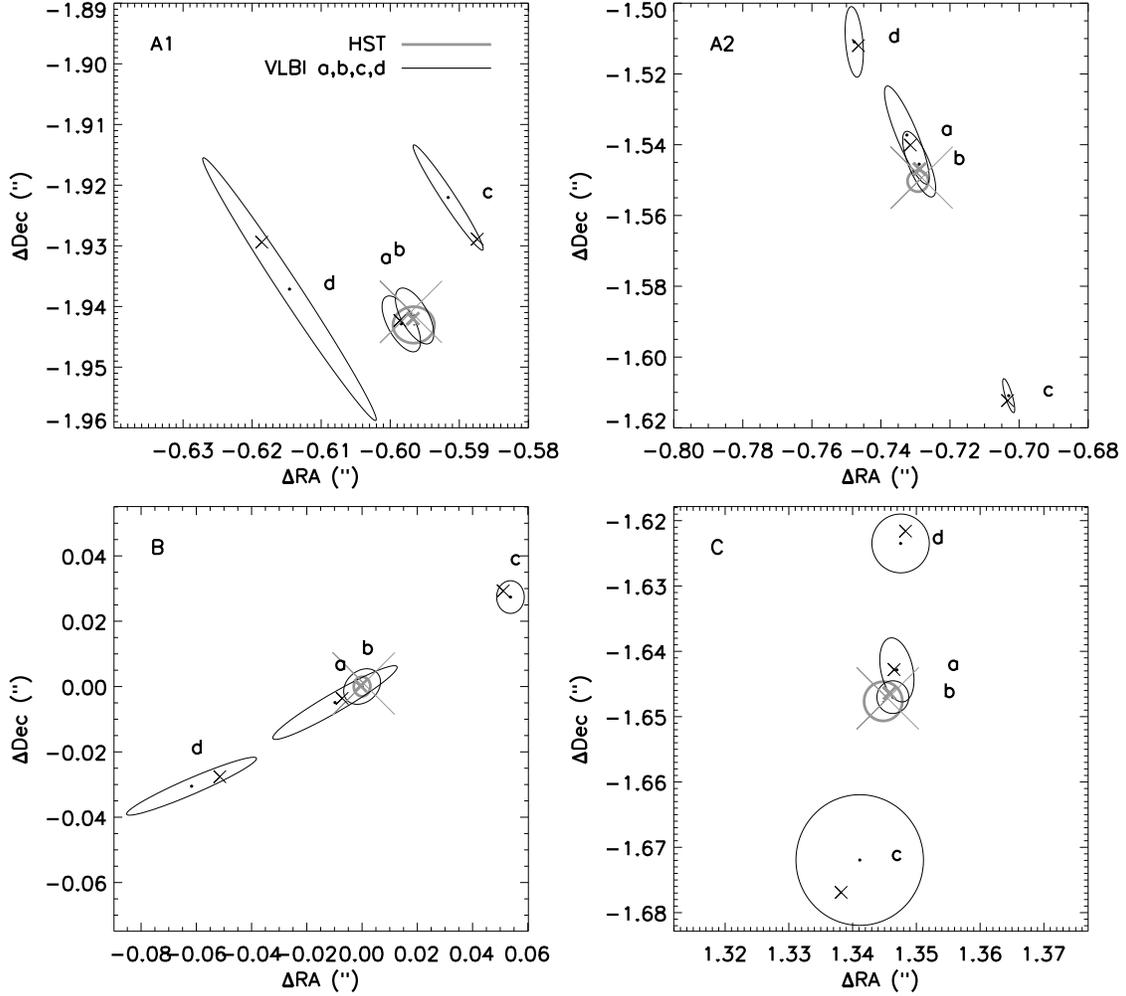}
  \caption{The constraints on the positions of the five sources used in our final
  model are shown in the image plane as error ellipses. Our estimates for the VLBI 
  core components ($a$ and $b$) and jet structures ($c$ and $d$) from \cite{ros00} are
  shown as black ellipses.  The error circles for the mid-IR positions are shown in gray.
  The best-fit positions for each component are indicated by the symbols and are 
  consistent with each error ellipse. The larger symbols correspond to the mid-IR 
  image positions, and the smaller, thick gray symbols correspond to VLBI component b. 
  }
\label{fig:errellipse}
\end{figure}

Fortunately, we can further test this model by adding the flux ratios and positions of the components detected in the radio with VLBI \citep{ros00,trot00}, where each quasar image is resolved into core and jet structures. 
We added the core components $a$ and $b$ detected in the VLBI maps, and the VLBI jet structures 
$c$ and $d$. The radio positions were constrained using the error ellipses shown in 
Figure~\ref{fig:errellipse},  which are estimated based on the outermost 
contours in Figures 1 and 2 in \citet{ros00}, and extend to the most distant ``clump'' in the jet structures.
For the $c$ and $d$ components in image C, we set the error ellipses to an arbitrarily large radius since 
they are not clearly resolved. Note that \citet{trot00} used tighter 
constraints on the component positions while ours are more conservative, especially along the jet 
structures, to allow more freedom in the component cross-identifications.  We find a better agreement 
($\Delta \chi^2 = 3$) between the optical and radio core positions when shifting the $HST$ images 
counter-clockwise by $0.1^{\circ}$, which is well within the $HST$ astrometric 
uncertainties\footnote{\scriptsize{See http://www.stsci.edu/hst/stis/strategies/pushing/documents/handbooks/currentIHB/c11\_datataking5.html}} 
of $\sim 0.5^{\circ}$.  
For the remaining models, we assume the rotated $HST$ positions for the mid-IR images (gray circles) 
and lens galaxies.  

We list the best-fit model parameters for a two--galaxy model in Table~\ref{tab:modelsMG0414}.  
Here, we adopt an SIS profile for G2 since it provides a marginally better fit in this case than 
a pseudo-Jaffe profile.  Our two--galaxy model has dof~$=41$, since there are four additional sources, 
each with three flux ratios and four pairs of $(x,y)$ image coordinates as constraints, and four 
additional source positions as free parameters.  
Our best-fit component positions are indicated with symbols in Figure~\ref{fig:errellipse}.  
There are systematic offsets between the VLBI centroids and best-fit source positions (see 
the $c$ and $d$ components in image A1), which are often much larger than the VLBI beam size of $2.55 \times 1.13$ mas.
However, these systematic offsets are along the direction of the physical extent of the jets and 
could easily be due to inaccurate cross-identification of the VLBI centroids and the systematic 
problems of VLBI maps in the presence of complex structures, which is why we adopted the error ellipses discussed earlier.  

\begin{figure}[h!]
  \centering
  \epsscale{1}
  \plottwo{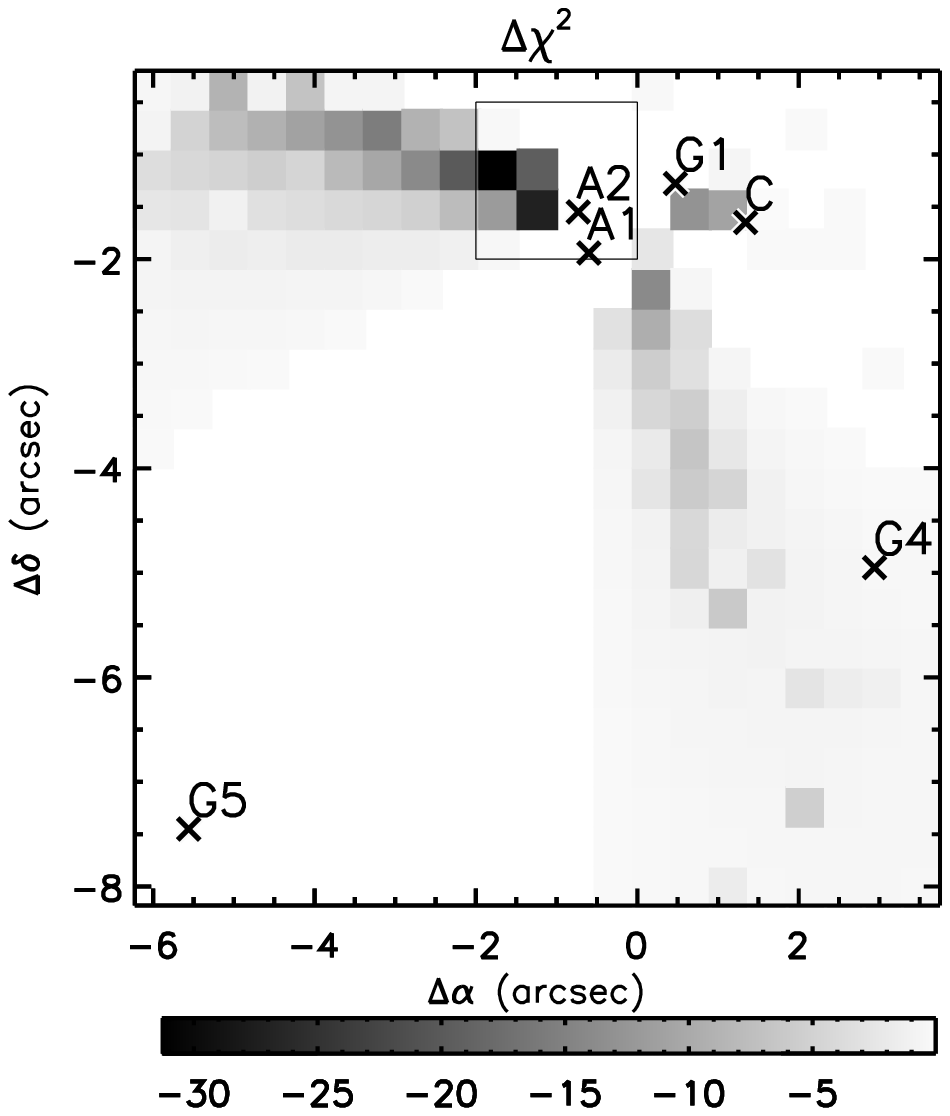}{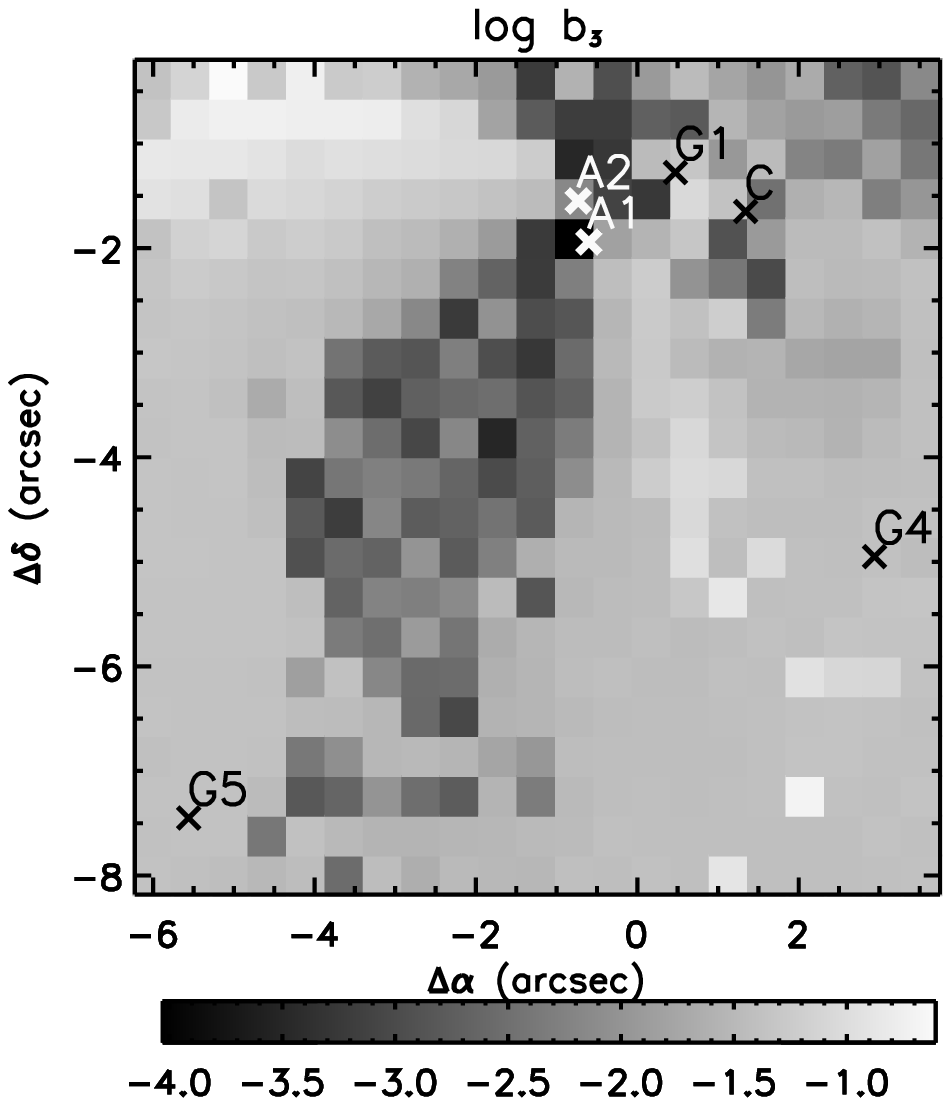}
  \plottwo{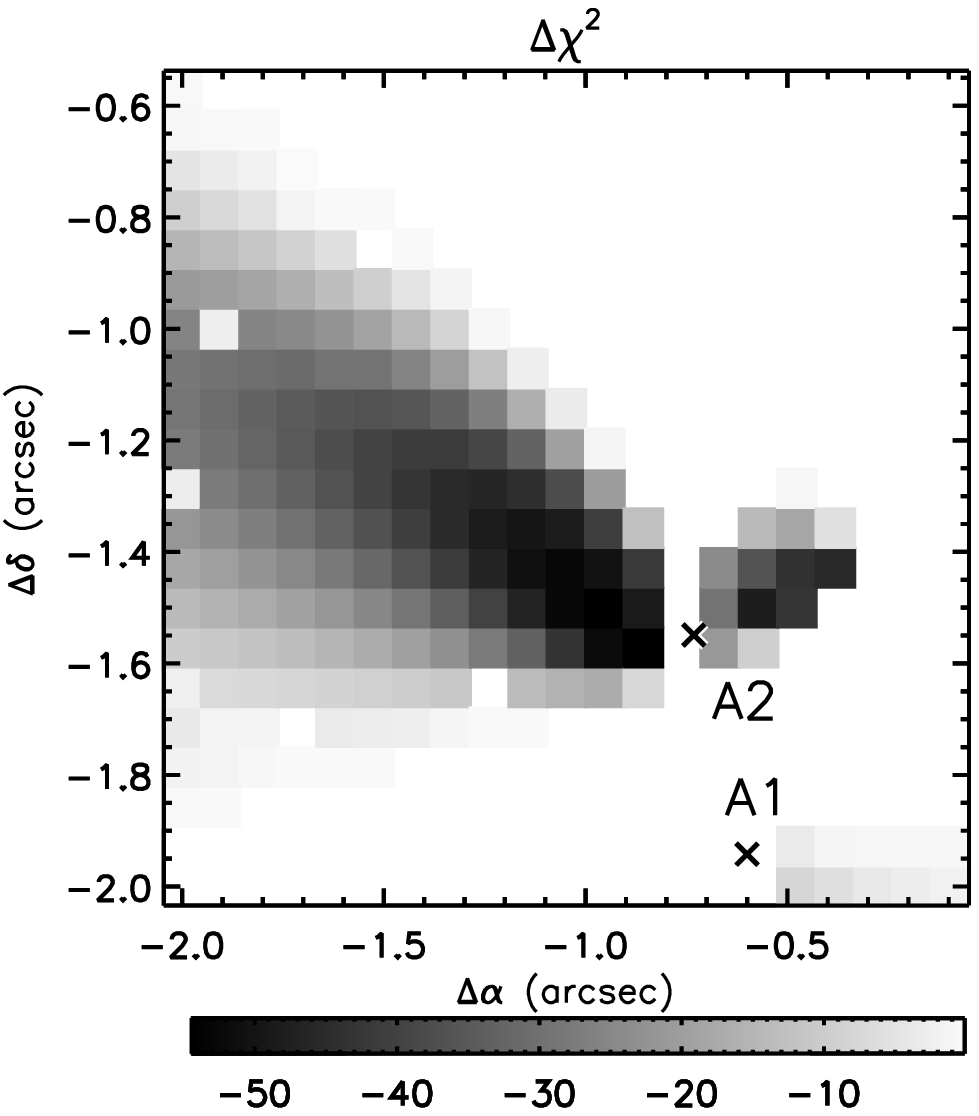}{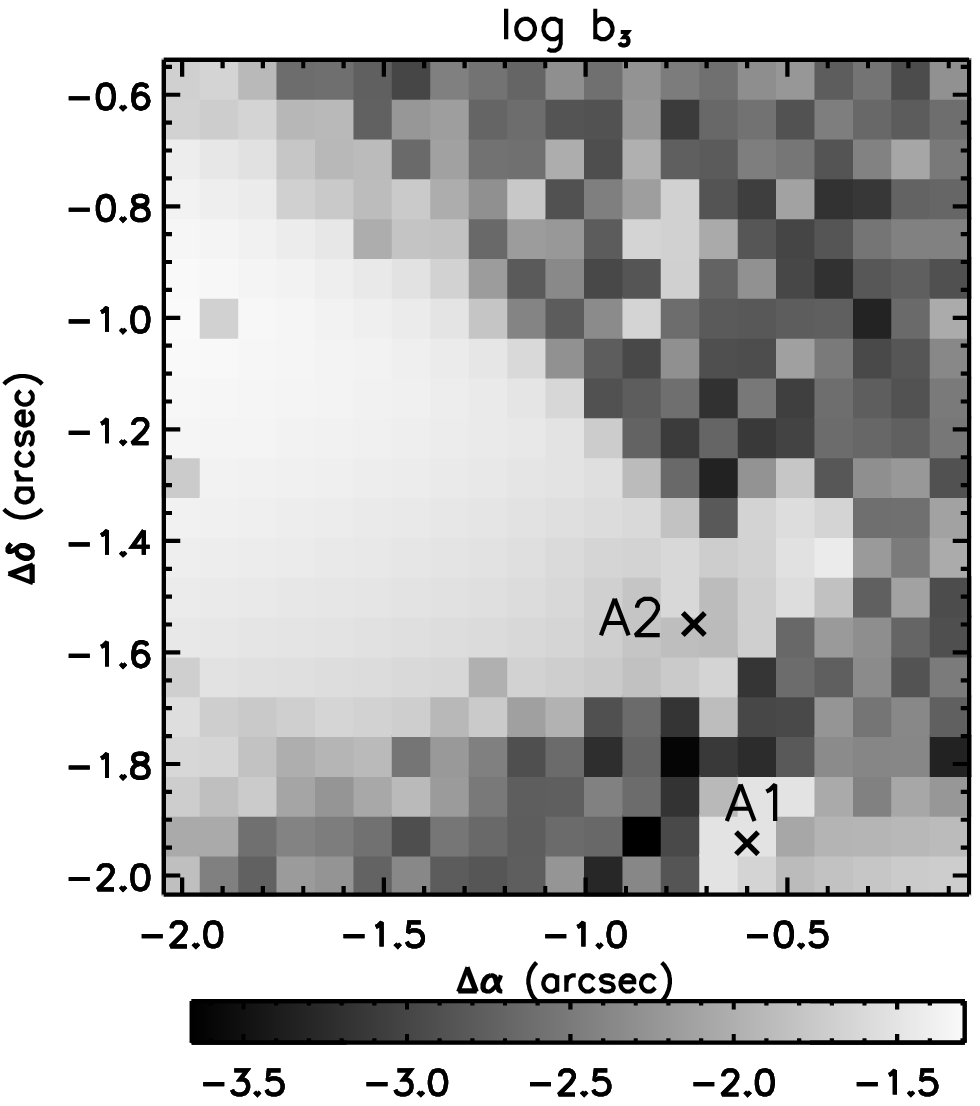}
  \caption{\footnotesize  
    The left panels show the $\Delta\chi^2$ improvement resulting from adding
    a third lens galaxy as a function of its 
    position, as in Fig.~\ref{fig:3gal}, but using both the mid-IR and radio 
    constraints on the models. The right panels show the best-fit critical radius for
    the substructure.  The bottom panels give an expanded view of the region 
    indicated by the box in the top-left panel. The positions of two nearby 
    galaxies G4 and G5 seen in the H-band image are also indicated.}
  \label{fig:3galr}
\end{figure}

Figure~\ref{fig:3galr} shows the resulting $\chi^2$ improvement when adding a third lens galaxy 
in the same fashion as before, but including the constraints from all four radio sources (a 
similar map using the radio constraints only is shown in the right panel of Figure~\ref{fig:3gal}). 
The best-fit mass scale $b_3$ is also shown. 
The overall results are consistent with those based only on the mid-IR data, but with a narrower 
region extending to the Northeast ($145^{\circ}$ North of West) for the position of the third galaxy.  
The parameters for the best three--galaxy model with $(x_3,y_3)=(-0\farcs97,-1\farcs39)$ and $\chi^2/$dof~$=1.6$ are listed 
in Table~\ref{tab:modelsMG0414}.  
We determined the $1\sigma$ error in the mass scale for G3 by varying $b_3$ in steps of 0\farcs0001 
around the best-fit value of $0\farcs007$ until the $\chi^2$ increased by one on either side of the minimum, while 
leaving the other lens parameters (including the external shear) allowed to vary.  Using this method, 
we obtained an error of 0\farcs001 for a third lens at this location. 
This corresponds to $10^{7.3\pm 0.2}\;M_\odot$ enclosed within the Einstein radius at $z_l=0.96$.  
The  statistical significance of this result is $>99\%$ based on the F-test.
We can also obtain a good fit with $b_3=0\farcs002$ at $(x_3,y_3)=(-0\farcs86,-1\farcs54)$ if the position of G3 is set initially closer to image A2.  
This degeneracy results because the model requires a larger mass for G3 to reproduce the observed  flux ratios when G3 is farther from image A2.  
Given this limitation, we can only constrain the mass scale to $0\farcs002< b_3 < 0\farcs008 $ based on the range of models in 
Figure~\ref{fig:3galr} that yield a comparable $\chi^2$ improvement, corresponding to masses within the Einstein radius of between $10^{6.2}$ and $10^{7.5}\;M_\odot$.
For simplicity, we adopt the final 3-galaxy model listed in Table 3 for the remainder of our analysis. 
  
\begin{figure}[h!]
	\begin{center}
          \epsscale{1}
        \plotone{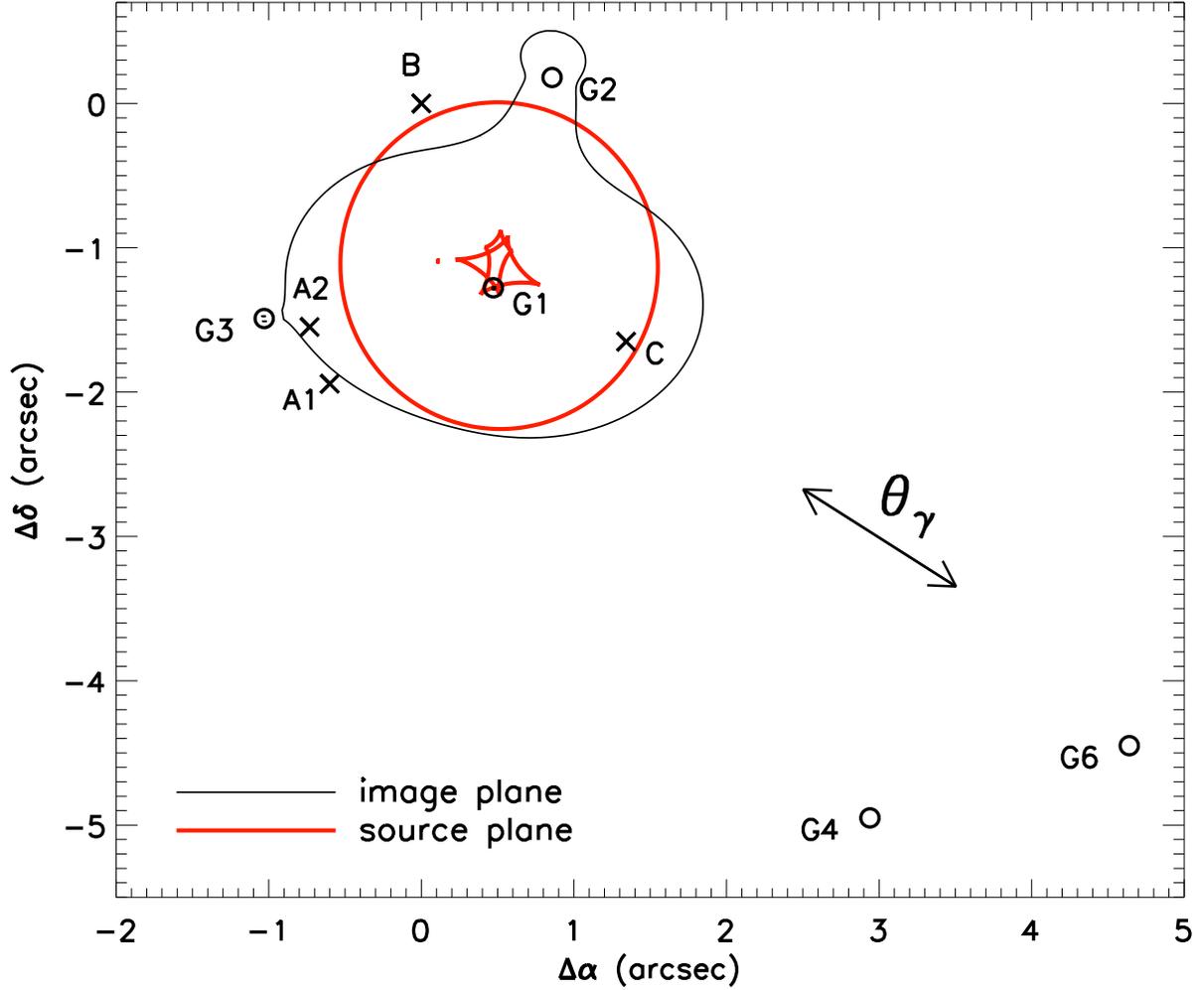}
	\end{center}
\caption{Best-fit critical (thin)
  and caustic (red) curves for our three--galaxy model when including both the mid-IR and
  radio constraints. Galaxy (image) positions are indicated with open circles (crosses). 
  The direction of the external shear is indicated with arrows, and points toward another galaxy (G6) detected in the H-band image.  }
\label{fig:CriticalCurves}
\end{figure}

The critical and caustic curves for our final three--galaxy model 
are shown in Figure~\ref{fig:CriticalCurves}.  The third galaxy distorts the critical curves near images A1 and A2, bringing their 
flux ratio back into agreement with the mid-IR data.   
G3 may be associated with a faint dwarf galaxy or a nonluminous dark matter substructure, 
as there are no nearby objects detected in the H-band image which are coincident with its predicted positions. 
Unfortunately, in the H-band image where we would best be able
to detect G3, the estimated position lies almost exactly on the
first Airy ring of image A2 (where the image subtraction residuals are high) 
and in a region with substantial emission from the quasar host galaxy,  
making it impossible to place useful limits on the mass-to-light ratio of G3.

The orientation of the external shear in our three--galaxy model, indicated 
in Figure~\ref{fig:CriticalCurves}, is in the same general direction as 
another object (G6) found 4\farcs4 Southwest of image C in the H-band image. 
As pointed out by \citet{fal97}, G6 is a possible compact group of galaxies,  
as it consists of three visible galaxies within 1'' of each other.
If we assume that G6 and G2 are at the same redshift as G1 and scale the Einstein radii 
of these objects by their I-band fluxes assuming a standard Faber-Jackson relation  
($L \propto \sigma^4$), we can estimate the deflection scale of G6 as 
 $b_6=b_{G2}(L_6/L_{G2})^{1/2}$, using the fact that $b\propto\sigma^2$ for an SIS. 
Adopting the I-band magnitudes of G2 (24.769) and G6 (23.16) from \cite{fal97}, 
we estimate $b_6 = 0\farcs34$.  The shear caused by G6 is then $\gamma_6=b_6/2r_6= 0.03$, 
where $r_6$ is the angular separation from G1, and the amplitude of any 
higher order perturbations is $b_6 b_{G1}/r_6^2 = 0.013$ \citep[see][]{koc06}. 
Since $\gamma_6<0.09$, G6 does not appear massive enough to fully explain the 
external shear in our three--galaxy model, although it likely contributes some shear.
The inclusion of G6 as a fourth lens mass yields a best-fit with $b_6=0\farcs21$, 
a reduction in shear amplitude (0.061), and a minor change in the shear orientation 
($53.3^{\circ}$) with $\chi^2/{\rm dof}=52.6/37=1.4$.

In an attempt to bring our reduced $\chi^2$ closer to unity, 
we tried adding a fifth lens galaxy on a grid of positions while optimizing the overall lens properties.  
We find the best improvement ($\Delta \chi^2\approx 7$) near image C, extending 
to a region $\sim$2'' North of G6.  However, the reduced $\chi^2$ does not improve ($46/34=1.4$).   
We conclude that additional substructures are unlikely to lead to a significant 
improvement of the fit given the constraints adopted here.  

We also checked the significance of G2 in our models since the redshift for G2 has not been 
measured, and there is a slight chance it lies at a different redshift than G1 and has a smaller gravitational influence 
than expected. We excluded G2 from our final 3-galaxy model and attempted to fit the available mid-IR and VLBI 
data using just G1 and G3, allowing the mass scales and external shear to vary along with the position and ellipticity 
of G1.  The best fit to the observed lens astrometry was significantly 
worse with a total $\chi^2$ of 470, which includes a factor of 10 increase in the image 
position $\chi^2$, a factor of 266 increase in the contribution from the lens (G1) position, 
and a factor of 3 increase in the contribution from the flux ratios.  Therefore, we confirm 
the earlier result that G2 is essential for reproducing the lens astrometry. 
The main consequence of G2 lying at a different redshift is that we would incorrectly convert its Einstein radius into a mass.

\clearpage

\section{Discussion}
\label{sec:disc}

Gravitationally lensed quasars are useful astrophysical tools because they allow us to study 
both the lens mass distribution and the background quasar.  Unlike microlensing, where time 
variability can be used to isolate the effect, the effects of millilensing need to be 
disentangled from other physics that modify flux ratios.  Mid-IR observations are ideal for 
this purpose because the effects of both microlensing and extinction are negligible or small \citep{mar12,slu13}.
The effect of microlensing of the dust torus on our observations can be estimated by 
comparing the average Einstein radius $R_E$ of stars in the lens galaxy to the dust 
sublimation radius $R_{min}$.  For a bolometric luminosity of roughly $1.4\times10^{47}$~ergs s$^{-1}$, 
estimated from correcting the total flux in our observations for a total magnification of 
38.5 and assuming no extinction, we find a minimum dust radius of $R_{min} \sim 5$~pc.
This is large compared to the average Einstein radius of $R_E\simeq0.0066$~pc, assuming a 
typical microlens mass of 0.3~M$_{\odot}$, and thus the microlensing effects should be no larger than 0.003~mag \citep{ref91}.    
In \citet{mac09}, we used the mid-IR flux ratios for the lens H1413+117 to show that flux ratio anomalies can detect companion lens galaxies, albeit a luminous companion in that system.  This technique can also be used for detecting dark subhaloes in the lens, down  to $10^5 M_{\odot}$ \citep{mao98,met02}.  As shown here and in \citet{min09},  the quadruply lensed quasar MG0414+0534 exhibits a flux ratio anomaly at mid-IR wavelengths that suggests there is millilensing substructure in the lens or along the line of sight to the quasar.   

When we model either the mid-IR or radio data or both using only two lens galaxies (the 
primary lens G1 and a secondary lens G2/object X), we are unable to reproduce the observed 
flux ratios. When we add a third lens G3 near image A2, we obtain a significantly improved 
fit to the flux ratios.  The optimal location of the substructure,  roughly 0\farcs3 
to the Northeast of image A2 and 1\farcs5 East of G1, was determined by adding 
a third lens on a grid of positions in the lens plane and varying its mass along with the 
overall lens properties. We are unable to associate the substructure with any source in the H-band image from $HST$, which suggests the third galaxy is either very faint or nonluminous.  
The mass of the substructure within the Einstein radius is in the range $10^{6.2}$ to $10^{7.5}\;M_\odot$ when modeled as an SIS, depending on the precise location of the substructure, and our results are statistically significant at $>99$\% confidence based on the F test.  Our findings are generally consistent with the less detailed models of \citet{min09}. The anomaly in the mid-IR suggests the presence of a third lens galaxy which we assume is at the redshift of the main lens, although the structure could theoretically lie at some other redshift along the line of sight \citep{xu12,ino12}.  There is also evidence for some contribution to the flux ratios from the neighboring galaxy group (G6) in the H-band image.  

Besides being relevant to studying the background quasar, the search for subhaloes in the lens is also cosmologically important, because 
it offers the only means of identifying dark, low mass halos other than detection of $\gamma$-ray annihilation signals by Fermi.  Sufficiently massive structures can be detected through their perturbations of image positions as seen with object X here, or in MG2016+112 \citep{mey06} and B1938+666 \citep{veg12}.  Lower mass structures can only be detected through flux ratio anomalies, where the challenge is to obtain flux ratios unaffected by absorption or microlensing.  The rest frame mid-IR satisfies these requirements but is limited at present by either poor resolution, as in Spitzer observations, or poor sensitivity, as in ground-based measurements.  The LBTI mid-IR imaging interferometer \citep{hinz03} on the Large Binocular Telescope (LBT) is a promising near-term solution, and a complete survey will be trivial with the advent of the James Webb Space Telescope.  

\acknowledgments
We thank the reviewer for his/her valuable comments and suggestions that improved the manuscript. 
This work is based on observations obtained at the Gemini Observatory
(with Program ID GN-2005B-Q-43), which is operated by the
Association of Universities for Research in Astronomy, Inc., under a cooperative agreement
with the NSF on behalf of the Gemini partnership: the National Science Foundation (United
States), the Science and Technology Facilities Council (United Kingdom), the
National Research Council (Canada), CONICYT (Chile), the Australian Research Council
(Australia), MinistÈrio da CiÍncia e Tecnologia (Brazil) 
and Ministerio de Ciencia, TecnologÌa e InnovaciÛn Productiva
(Argentina). This work is based in part on observations made with the NASA/ESA Hubble 
Space Telescope. Support for program GO-7495 was provided by NASA 
through a grant from the Space Telescope Science Institute, which is
operated by AURA, Inc., under NASA contract NAS5-2655.  EA is
partially supported by National Science Foundation CAREER Grant No.
0645416. CSK is supported by NSF grant AST-1009756.

{\it Facilities:} \facility{Gemini (Michelle)}, \facility{HST (NICMOS)}.

\bibliography{refs}

\begin{thebibliography}{48}
\expandafter\ifx\csname natexlab\endcsname\relax\def\natexlab#1{#1}\fi

\bibitem[{{Agol} {et~al.}(2009){Agol}, {Gogarten}, {Gorjian}, \&
  {Kimball}}]{ago09}
{Agol}, E., {Gogarten}, S.~M., {Gorjian}, V., \& {Kimball}, A. 2009, \apj, 697,
  1010

\bibitem[{{Bate} {et~al.}(2008){Bate}, {Floyd}, {Webster}, \& {Wyithe}}]{bat08}
{Bate}, N.~F., {Floyd}, D.~J.~E., {Webster}, R.~L., \& {Wyithe}, J.~S.~B. 2008,
  \mnras, 391, 1955

\bibitem[{{Bate} {et~al.}(2011){Bate}, {Floyd}, {Webster}, \& {Wyithe}}]{bat11}
---. 2011, \apj, 731, 71

\bibitem[{{Belokurov} {et~al.}(2007){Belokurov}, {Zucker}, {Evans}, {Kleyna},
  {Koposov}, {Hodgkin}, {Irwin}, {Gilmore}, {Wilkinson}, {Fellhauer},
  {Bramich}, {Hewett}, {Vidrih}, {De Jong}, {Smith}, {Rix}, {Bell}, {Wyse},
  {Newberg}, {Mayeur}, {Yanny}, {Rockosi}, {Gnedin}, {Schneider}, {Beers},
  {Barentine}, {Brewington}, {Brinkmann}, {Harvanek}, {Kleinman}, {Krzesinski},
  {Long}, {Nitta}, \& {Snedden}}]{bel07}
{Belokurov}, V., {Zucker}, D.~B., {Evans}, N.~W., {et~al.} 2007, \apj, 654, 897

\bibitem[{{Bullock} {et~al.}(2000){Bullock}, {Kravtsov}, \& {Weinberg}}]{bul00}
{Bullock}, J.~S., {Kravtsov}, A.~V., \& {Weinberg}, D.~H. 2000, \apj, 539, 517

\bibitem[{{Cohen} {et~al.}(1999){Cohen}, {Walker}, {Carter}, {Hammersley},
  {Kidger}, \& {Noguchi}}]{coh99}
{Cohen}, M., {Walker}, R.~G., {Carter}, B., {et~al.} 1999, \aj, 117, 1864

\bibitem[{{Congdon} \& {Keeton}(2005)}]{con05}
{Congdon}, A.~B., \& {Keeton}, C.~R. 2005, \mnras, 364, 1459

\bibitem[{{Dalal} \& {Kochanek}(2002)}]{dal02}
{Dalal}, N., \& {Kochanek}, C.~S. 2002, \apj, 572, 25

\bibitem[{{Evans} \& {Witt}(2003)}]{eva03}
{Evans}, N.~W., \& {Witt}, H.~J. 2003, \mnras, 345, 1351

\bibitem[{{Fadely} \& {Keeton}(2012)}]{fad12}
{Fadely}, R., \& {Keeton}, C.~R. 2012, \mnras, 419, 936

\bibitem[{{Falco} {et~al.}(1997){Falco}, {Lehar}, \& {Shapiro}}]{fal97}
{Falco}, E.~E., {Lehar}, J., \& {Shapiro}, I.~I. 1997, \aj, 113, 540

\bibitem[{{Hewitt} {et~al.}(1992){Hewitt}, {Turner}, {Lawrence}, {Schneider},
  \& {Brody}}]{hew92}
{Hewitt}, J.~N., {Turner}, E.~L., {Lawrence}, C.~R., {Schneider}, D.~P., \&
  {Brody}, J.~P. 1992, \aj, 104, 968

\bibitem[{{Hinz} {et~al.}(2003){Hinz}, {Angel}, {McCarthy}, {Hoffman}, \&
  {Peng}}]{hinz03}
{Hinz}, P.~M., {Angel}, J.~R.~P., {McCarthy}, Jr., D.~W., {Hoffman}, W.~F., \&
  {Peng}, C.~Y. 2003, in Society of Photo-Optical Instrumentation Engineers
  (SPIE) Conference Series, Vol. 4838, Society of Photo-Optical Instrumentation
  Engineers (SPIE) Conference Series, ed. W.~A. {Traub}, 108--112

\bibitem[{{Inoue} \& {Takahashi}(2012)}]{ino12}
{Inoue}, K.~T., \& {Takahashi}, R. 2012, ArXiv e-prints

\bibitem[{{Keeton}(2001)}]{kee01}
{Keeton}, C.~R. 2001, ArXiv Astrophysics e-prints

\bibitem[{{Keeton} {et~al.}(2005){Keeton}, {Gaudi}, \& {Petters}}]{kee05}
{Keeton}, C.~R., {Gaudi}, B.~S., \& {Petters}, A.~O. 2005, \apj, 635, 35

\bibitem[{{Keeton} {et~al.}(1997){Keeton}, {Kochanek}, \& {Seljak}}]{kee97}
{Keeton}, C.~R., {Kochanek}, C.~S., \& {Seljak}, U. 1997, \apj, 482, 604

\bibitem[{{Keeton} \& {Moustakas}(2009)}]{kee09}
{Keeton}, C.~R., \& {Moustakas}, L.~A. 2009, \apj, 699, 1720

\bibitem[{{Klypin} {et~al.}(1999){Klypin}, {Kravtsov}, {Valenzuela}, \&
  {Prada}}]{kly99}
{Klypin}, A., {Kravtsov}, A.~V., {Valenzuela}, O., \& {Prada}, F. 1999, \apj,
  522, 82

\bibitem[{{Kochanek}(2006)}]{mey06}
{Kochanek}, C.~S. 2006, in Saas-Fee Advanced Course 33: Gravitational Lensing:
  Strong, Weak and Micro, ed. G.~{Meylan}, P.~{Jetzer}, P.~{North},
  P.~{Schneider}, C.~S. {Kochanek}, \& J.~{Wambsganss}, 91--268

\bibitem[{{Kochanek} \& {Dalal}(2004)}]{koc04}
{Kochanek}, C.~S., \& {Dalal}, N. 2004, \apj, 610, 69

\bibitem[{{Kochanek} {et~al.}(2006){Kochanek}, {Morgan}, {Falco}, {McLeod},
  {Winn}, {Dembicky}, \& {Ketzeback}}]{koc06}
{Kochanek}, C.~S., {Morgan}, N.~D., {Falco}, E.~E., {et~al.} 2006, \apj, 640,
  47

\bibitem[{{Koopmans} {et~al.}(2009){Koopmans}, {Bolton}, {Treu}, {Czoske},
  {Auger}, {Barnab{\`e}}, {Vegetti}, {Gavazzi}, {Moustakas}, \&
  {Burles}}]{koo09}
{Koopmans}, L.~V.~E., {Bolton}, A., {Treu}, T., {et~al.} 2009, \apjl, 703, L51

\bibitem[{{Kratzer} {et~al.}(2011){Kratzer}, {Richards}, {Goldberg}, {Oguri},
  {Kochanek}, {Hodge}, {Becker}, \& {Inada}}]{kra11}
{Kratzer}, R.~M., {Richards}, G.~T., {Goldberg}, D.~M., {et~al.} 2011, \apjl,
  728, L18

\bibitem[{{Lawrence} {et~al.}(1995){Lawrence}, {Elston}, {Januzzi}, \&
  {Turner}}]{law95}
{Lawrence}, C.~R., {Elston}, R., {Januzzi}, B.~T., \& {Turner}, E.~L. 1995,
  \aj, 110, 2570

\bibitem[{{MacLeod} {et~al.}(2009){MacLeod}, {Kochanek}, \& {Agol}}]{mac09}
{MacLeod}, C.~L., {Kochanek}, C.~S., \& {Agol}, E. 2009, \apj, 699, 1578

\bibitem[{{Mao} \& {Schneider}(1998)}]{mao98}
{Mao}, S., \& {Schneider}, P. 1998, \mnras, 295, 587

\bibitem[{{Metcalf} \& {Zhao}(2002)}]{met02}
{Metcalf}, R.~B., \& {Zhao}, H. 2002, \apjl, 567, L5

\bibitem[{{Minezaki} {et~al.}(2009){Minezaki}, {Chiba}, {Kashikawa}, {Inoue},
  \& {Kataza}}]{min09}
{Minezaki}, T., {Chiba}, M., {Kashikawa}, N., {Inoue}, K.~T., \& {Kataza}, H.
  2009, \apj, 697, 610

\bibitem[{{Moore} {et~al.}(1999){Moore}, {Ghigna}, {Governato}, {Lake},
  {Quinn}, {Stadel}, \& {Tozzi}}]{moo99}
{Moore}, B., {Ghigna}, S., {Governato}, F., {et~al.} 1999, \apjl, 524, L19

\bibitem[{{Mu{\~n}oz} {et~al.}(2011){Mu{\~n}oz}, {Mediavilla}, {Kochanek},
  {Falco}, \& {Mosquera}}]{mun11}
{Mu{\~n}oz}, J.~A., {Mediavilla}, E., {Kochanek}, C.~S., {Falco}, E.~E., \&
  {Mosquera}, A.~M. 2011, \apj, 742, 67

\bibitem[{{Pooley} {et~al.}(2012){Pooley}, {Rappaport}, {Blackburne},
  {Schechter}, \& {Wambsganss}}]{poo12}
{Pooley}, D., {Rappaport}, S., {Blackburne}, J.~A., {Schechter}, P.~L., \&
  {Wambsganss}, J. 2012, \apj, 744, 111

\bibitem[{{Refsdal} \& {Stabell}(1991)}]{ref91}
{Refsdal}, S., \& {Stabell}, R. 1991, \aap, 250, 62

\bibitem[{{Roche}(2004)}]{roc04}
{Roche}, P.~F. 2004, Advances in Space Research, 34, 583

\bibitem[{{Ros} {et~al.}(2000){Ros}, {Guirado}, {Marcaide}, {P{\'e}rez-Torres},
  {Falco}, {Mu{\~n}oz}, {Alberdi}, \& {Lara}}]{ros00}
{Ros}, E., {Guirado}, J.~C., {Marcaide}, J.~M., {et~al.} 2000, \aap, 362, 845

\bibitem[{{Schechter} \& {Moore}(1993)}]{sch93}
{Schechter}, P.~L., \& {Moore}, C.~B. 1993, \aj, 105, 1

\bibitem[{{Schechter} \& {Wambsganss}(2002)}]{sch02}
{Schechter}, P.~L., \& {Wambsganss}, J. 2002, \apj, 580, 685

\bibitem[{{Sluse} {et~al.}(2013){Sluse}, {Kishimoto}, {Anguita}, {Wucknitz}, \&
  {Wambsganss}}]{slu13}
{Sluse}, D., {Kishimoto}, M., {Anguita}, T., {Wucknitz}, O., \& {Wambsganss},
  J. 2013, \aap, 553, A53

\bibitem[{{Stalevski} {et~al.}(2012){Stalevski}, {Jovanovi{\'c}},
  {Popovi{\'c}}, \& {Baes}}]{mar12}
{Stalevski}, M., {Jovanovi{\'c}}, P., {Popovi{\'c}}, L.~{\v C}., \& {Baes}, M.
  2012, \mnras, 425, 1576

\bibitem[{{Strigari} {et~al.}(2008){Strigari}, {Koushiappas}, {Bullock},
  {Kaplinghat}, {Simon}, {Geha}, \& {Willman}}]{str08}
{Strigari}, L.~E., {Koushiappas}, S.~M., {Bullock}, J.~S., {et~al.} 2008, \apj,
  678, 614

\bibitem[{{Tonry} \& {Kochanek}(1999)}]{ton99}
{Tonry}, J.~L., \& {Kochanek}, C.~S. 1999, \aj, 117, 2034

\bibitem[{{Trotter} {et~al.}(2000){Trotter}, {Winn}, \& {Hewitt}}]{trot00}
{Trotter}, C.~S., {Winn}, J.~N., \& {Hewitt}, J.~N. 2000, \apj, 535, 671

\bibitem[{{Tseliakhovich} \& {Hirata}(2010)}]{hirata}
{Tseliakhovich}, D., \& {Hirata}, C. 2010, \prd, 82, 083520

\bibitem[{{Vegetti} {et~al.}(2012){Vegetti}, {Lagattuta}, {McKean}, {Auger},
  {Fassnacht}, \& {Koopmans}}]{veg12}
{Vegetti}, S., {Lagattuta}, D.~J., {McKean}, J.~P., {et~al.} 2012, \nat, 481,
  341

\bibitem[{{Willman} {et~al.}(2005){Willman}, {Blanton}, {West}, {Dalcanton},
  {Hogg}, {Schneider}, {Wherry}, {Yanny}, \& {Brinkmann}}]{wil05}
{Willman}, B., {Blanton}, M.~R., {West}, A.~A., {et~al.} 2005, \aj, 129, 2692

\bibitem[{{Xu} {et~al.}(2012){Xu}, {Mao}, {Cooper}, {Gao}, {Frenk}, {Angulo},
  \& {Helly}}]{xu12}
{Xu}, D.~D., {Mao}, S., {Cooper}, A.~P., {et~al.} 2012, \mnras, 421, 2553

\bibitem[{{Yoo} {et~al.}(2006{\natexlab{a}}){Yoo}, {Kochanek}, {Falco}, \&
  {McLeod}}]{yoo06a}
{Yoo}, J., {Kochanek}, C.~S., {Falco}, E.~E., \& {McLeod}, B.~A.
  2006{\natexlab{a}}, \apj, 642, 22

\bibitem[{{Yoo} {et~al.}(2006{\natexlab{b}}){Yoo}, {Tinker}, {Weinberg},
  {Zheng}, {Katz}, \& {Dav{\'e}}}]{yoo06}
{Yoo}, J., {Tinker}, J.~L., {Weinberg}, D.~H., {et~al.} 2006{\natexlab{b}},
  \apj, 652, 26

\end{thebibliography}
\bibliographystyle{apj}

\end{document}